\documentclass[useAMS,usenatbib]{mn2e}
\usepackage{bm, amsmath, amssymb, graphicx}
\usepackage[english]{babel}
\title[The population of AM CVn stars]{The population of AM CVn stars from the Sloan Digital Sky Survey}
\author[G.\,H.\,A. Roelofs et al.]{G.\,H.\,A.~Roelofs,\thanks{E-mail: groelofs@cfa.harvard.edu} G.~Nelemans, and P.\,J.~Groot\\Department of Astrophysics, Radboud University, Toernooiveld 1, 6525 ED Nijmegen, The Netherlands}

\newcommand{\beq}{\begin{equation}}
\newcommand{\eeq}{\end{equation}}
\newcommand{\ud}{\mathrm{d}}

\begin{document}
\maketitle

\begin{abstract}
The AM Canum Venaticorum stars are rare interacting white dwarf binaries, whose formation and evolution are still poorly known. The Sloan Digital Sky Survey provides, for the first time, a sample of 6 AM CVn stars (out of a total population of 18) that is sufficiently homogeneous that we can start to study the population in some detail.

We use the Sloan sample to `calibrate' theoretical population synthesis models for the space density of AM CVn stars. We consider optimistic and pessimistic models for different theoretical formation channels, which yield predictions for the local space density that are more than two orders of magnitude apart. When calibrated with the observations, all models give a local space density $\rho_0=1-3\times10^{-6}$\,pc$^{-3}$, which is lower than expected.

We discuss the implications for the formation of AM CVn stars, and conclude that at least one of the dominant formation channels (the double-degenerate channel) has to be suppressed relative to the optimistic models. In the framework of the current models this suggests that the mass transfer between white dwarfs usually cannot be stabilized.
We furthermore discuss evolutionary effects that have so far not been considered in population synthesis models, but which could be of influence for the observed population.
We finish by remarking that, with our lower space density, the expected number of Galactic AM CVn stars resolvable by gravitational-wave detectors like \emph{LISA} should be lowered from current estimates, to about $1,000$ for a mission duration of one year.
\end{abstract}


\section{Introduction}

The Sloan Digital Sky Survey (SDSS; \citealt{york}) is the largest photometric and spectroscopic sky survey ever carried out, consisting of imaging in five broad-band filters ($u,g,r,i,z$) and follow-up spectroscopy of a large fraction of targets outside the stellar (main-sequence) locus. The first phase, SDSS-I (i.e.\ Data Release~5, \citealt{adelman}) contains photometry of 8000 square degrees concentrated on the northern Galactic cap, down to a limiting magnitude of about $g=22$, and spectroscopy of about 70\% of that area (we shall refer to these as the `photometric database' and the `spectroscopic database', and we will use the dereddened magnitudes according to \citet{schlegel} throughout this paper). The spectroscopic database contains about 1 million objects, which have been obtained in about 1700 pointings (`tiles') using a 640-fibre multi-object spectrograph. Although the spectroscopic follow-up to the imaging is mainly tailored towards finding quasars and galaxies, Galactic sources off the main sequence also make their way into the spectroscopic sample, serendipitously or sometimes targeted as white dwarfs (WDs) and hot standard stars.

Among the serendipitous targets are AM Canum Venaticorum stars: ultracompact binaries that consist of a white dwarf accreting helium-rich matter from a (semi-)degenerate donor star. They are most easily picked up as helium-emission-line objects in the SDSS spectroscopic database. To date, six such systems have been found in the SDSS (\citealt{roelofs,anderson,groot06}), a significant increase in the number of systems -- only 10 AM CVn stars were known prior to the SDSS, and another two were recently discovered as supernova candidates \citep{ww05,prieto}. The spectroscopic sample of emission-line systems from the SDSS is the first that is sufficiently large, complete, and homogeneous that a study of the AM CVn population can be attempted. Since it is still a small sample, the approach in this paper shall be to combine this observed sample with theoretical predictions for the AM CVn stars \citep{nelemans,nelemans2004}, in order to `calibrate' the model populations.

This will hopefully give us more insight into the as yet uncertain formation and evolution of these binaries. In addition, the AM CVn stars are of particular interest for gravitational-wave astrophysics: they are the only \emph{known} objects that can be detected with space-borne gravitational-wave detectors such as the proposed \emph{Laser Interferometer Space Antenna (LISA)} mission \citep{stroeer2005,stroeer2006,roelofshst}. Depending on their space density, as many as $\sim$11,000 AM CVn stars can be resolved throughout the Galaxy by a \emph{LISA}-type instrument \citep{nelemans2004}.

\section{Calibrating population synthesis models for the AM CVn stars}

\subsection{Formation channels}

Three formation channels have been proposed to contribute to the population of AM CVn stars. In the first of these, the `WD channel', a close pair of detached white dwarfs is formed via two common-envelope phases, which is subsequently driven towards Roche-lobe overflow by gravitational-wave emission (e.g.\ \citealt{nather,ty,nelemans}). After the onset of Roche-lobe overflow at an orbital period of a few minutes, the system evolves towards longer orbital periods as the degenerate donor star expands upon mass loss.

In the second scenario, the `He-star channel', a white dwarf accretes from an initially non-degenerate helium star, while the binary evolves to shorter orbital periods. The helium burning in the core of the donor star quenches as a result of the mass loss and the system reaches an orbital period minimum around $\sim$10 minutes, at which point the donor star has become semi-degenerate and the system starts to evolve towards longer orbital periods again \citep{tf,ibentutukov,nelemans}.

The third, `evolved-CV channel' was suggested more recently by \citet{podsi} and involves a Cataclysmic Variable (CV) in which the mass transfer starts by the time the donor star starts to evolve off the hydrogen main sequence. Depending on the level of hydrogen depletion that is reached, such a system may shift its orbital period minimum from about one hour (the normal orbital period minimum for degenerate hydrogen-rich matter) down to the orbital period minimum for helium-rich matter at $\sim$10 minutes, with most such systems reaching intermediate orbital period minima due to remaining fractions of hydrogen in their cores.

The objects emerging from each of these channels are expected to look like helium-rich, ultracompact AM CVn stars that evolve towards longer orbital periods and lower mass transfer rates, driven by angular momentum losses due to the emission of gravitational waves.

\subsection{Recent observational \& theoretical results}
\label{recent}

\citet{bildsten} have recently shown that, at long orbital periods, the AM CVn stars are expected to be dominated in the optical by an accretor that cools as if it were a stand-alone white dwarf. This is due to the rapidly-decreasing mass accretion rate losing control over the thermal state of the accretor, and the accretion luminosity dropping below the luminosity of the accretor. The steep decrease in mass transfer rate with orbital period, which is a consequence of the gravitational-wave losses being a steep function of orbital period, should cause the transition from disc-dominated to accretor-dominated spectra to occur over a fairly small range in orbital period. Observationally, this transition from a disc-dominated optical spectrum to an accretor-dominated spectrum is seen to occur at orbital periods of about 30 minutes, with systems at longer orbital periods spending most of their time in a `low' state characterized by helium emission lines from the disc and a continuum that is apparently dominated by the photosphere of the accretor. Systems at shorter orbital periods spend most of their time in a `high' state characterized by an absorption-line spectrum dominated by the accretion disc. Two recent additions to the AM CVn family, SDSS J1240 \citep{roelofs} and `SN\,2003aw' \citep{ww03,nogami,roelofsaw} at $P=37$ and 34 minutes respectively, appear to spend most of their time in an emission-line state dominated by the accretor's photosphere, going into outburst not more than about once a year. Shorter-period systems such as CR Boo and V803 Cen at $P=25$ and 27 minutes, on the other hand, spend most of their time in a high(er) state characterized by an accretion-disc-dominated absorption-line spectrum; they are seen to spend less than 10 per cent of their time in a low, emission-line state \citep{patterson97,patterson00}.

Assuming that, statistically, there is a transition from disc-dominated to accretor-dominated spectra at $P_\mathrm{min}=30$ minutes, and knowing the thermal evolution of the accretor-dominated systems, we can parametrize the colours and absolute magnitudes of these `long-period' AM CVns at $P>30$\,min with their orbital period. This allows us to determine the completeness of spectroscopic follow-up in the SDSS as a function of orbital period, by measuring the completeness as a function of colour. If we further assume a certain orbital period distribution for the AM CVn stars based on population synthesis studies, we can convolve this distribution with the completeness as a function of orbital period to obtain the total number of emission-line AM CVn stars present in the SDSS-I, from which we can derive the surface and space densities of these stars.

Population synthesis studies of AM CVn stars have yielded predictions about their space density, but these are thought to be uncertain to two orders of magnitude \citep{nelemans,nelemans2004}. Interestingly, the distribution of orbital periods is relatively well-known; it is mainly the absolute number of systems that is uncertain. One of the important reasons for that is the fact that the angular momentum loss due to gravitational waves, which directly drives the evolution towards longer orbital periods in these systems, is a steep function of the orbital period. This causes the fraction of systems at long orbital periods ($P>30$\,min), for instance, to be very similar for different population synthesis models since the evolution from the start of Roche-lobe overflow to this long-period regime is relatively instantaneous -- it takes only about $10^8$ years. For all model populations as described in \citet{nelemans,nelemans2004} we find only a $\leq2\%$ fraction of systems at short orbital periods ($P<30$\,min).

This allows us to calibrate the total space density of AM CVn stars based on the emission-line population in the SDSS and the predicted orbital period distributions from population synthesis studies. To assess the influence of the exact shape of the orbital period distribution, we can calculate the total space density for different model populations and see how they differ.

\subsection{Formal method}

We wish to determine the space density of emission-line AM CVn stars from the observed $N_\mathrm{spec}=6$ systems in the spectroscopic database of SDSS-I. To this end we simulate observations of model populations to see how many AM CVn stars we would have found in the SDSS-I if this population corresponded to the true population. The expected number of AM CVns in the spectroscopic database is given by an integral over the sampled volume, multiplied by the modelled space density $\rho'$ and by the completeness of spectroscopic follow-up $C$. The efficiency for identifying emission-line AM CVn stars once they have made their way into Sloan's spectroscopic database is assumed to be (nearly) 100\%: their characteristic helium-emission-line spectra are easily picked up using a combination of automated searches and visual inspection of spectra, as done by us and independently by \citet{anderson}. The completeness of spectroscopic follow-up, i.e.\ the fraction of AM CVn stars in the photometric database of which a spectrum was taken, is expected to be far from 100\%, and in addition it is expected to be a function of colour and apparent magnitude. One of our main concerns shall therefore be to determine this completeness.

We express the expected number $N'_\mathrm{spec}$ of emission-line AM CVn stars in the spectroscopic database of the SDSS, for a given model population, as an integral over the orbital period $P$ and the distance $d$, while summing over all spectroscopic pointings (primed quantities indicate model predictions):
\begin{equation}
N'_\mathrm{spec} =
\sum_{\mathrm{tiles}}\, \int\limits_{P_\mathrm{min}}^{\infty}\!\!\ud P\!\!\!\int\limits_{0}^{d_\mathrm{max}(P)}\!\!\!\!\!\ud d\,\,\Omega_\mathrm{tile} d^2\,\, C(P,b_\mathrm{tile},d)\,\,\rho '(P,b_\mathrm{tile},d)
\label{Nprime}
\end{equation}
Here, $C(P,b_\mathrm{tile},d)$ is the completeness of spectroscopic follow-up as a function of orbital period $P$ (which we link to effective temperature and thus colour via a certain prescription), as a function of distance $d$ (linked to apparent magnitude $g$ via a certain prescription), and as a function of Galactic latitude $b$. We determine this completeness \emph{a posteriori} from the Sloan databases by counting the fractions of objects that have spectroscopic follow-up in colour, magnitude and Galactic latitude intervals. The quantity $\rho'(P,b_\mathrm{tile},d)$ is the population synthesis prediction for the space density of systems as a function of orbital period, Galactic latitude and distance. The distance $d_\mathrm{max}(P)$ is the distance at which an object of orbital period $P$ would have an apparent magnitude $g_\mathrm{max}$, where $g_\mathrm{max}$ is a suitably chosen magnitude limit. Finally, $\Omega_\mathrm{tile}$ is the solid angle covered by each spectroscopic tile (pointing), over which we take the sum to get the total area covered by the spectroscopic database. Thus the integral $\int\!\Omega_\mathrm{tile}d^2\ud d$ expresses the volume corresponding to a pointing. The solid angle covered by each pointing is slightly variable, but we will simply use an effective solid angle $\Omega_\mathrm{tile}=\Omega_\mathrm{spec}/N_\mathrm{tiles}$, where $\Omega_\mathrm{spec}=5700$ square degrees is the total solid angle covered by the spectroscopic database, and $N_\mathrm{tiles}=1700$ is the number of spectroscopic tiles (pointings) over which we take the sum.

The observed local space density $\rho_0$ at Galactic height $z=0$, based on the observed $N_\mathrm{spec}=6$, is then given by
\beq
\rho_0 = \frac{N_\mathrm{spec}}{N'_\mathrm{spec}}\rho'_0
\label{transform}
\eeq
where $\rho'_0$ is the local space density of the model population.

The total number of emission-line AM CVn stars in the photometric database of SDSS-I down to $g=g_\mathrm{max}$ is obtained by setting $C(P,b_\mathrm{tile},d)=1$ in eq.\ \ref{Nprime} and multiplying by $\Omega_\mathrm{phot}/\Omega_\mathrm{spec}$, where $\Omega_\mathrm{phot}=8000$ square degrees is the solid angle on the celestial sphere covered by the photometric database:
\begin{equation}
N_\mathrm{phot} =
\frac{N_\mathrm{spec}}{N'_\mathrm{spec}}\frac{\Omega_\mathrm{phot}}{N_\mathrm{tiles}}\,\sum_{\mathrm{tiles}}\, \int\limits_{P_\mathrm{min}}^{\infty}\!\!\ud P\!\!\!\int\limits_{0}^{d_\mathrm{max}(P)}\!\!\!\!\!d^2\,\ud d
\,\, \rho '(P,b_\mathrm{tile},d)
\label{Nphot}
\end{equation}

This value $N_\mathrm{phot}$ divided by the area covered by the photometric database $\Omega_\mathrm{phot}$ gives the surface density on the sky $\sigma$ of emission-line AM CVn stars down to $g=g_\mathrm{max}$ at the typical Galactic latitudes covered by the SDSS-I.

\subsection{Population synthesis models}
\label{models}

The double-degenerate WD channel and the single-degenerate He-star channel are thought to be the dominant contributors to the AM CVn population, with the evolved-CV channel contributing at the $<2\%$ level \citep{nelemans2004}. As in \citet{nelemans} we model two scenarios for the total population of AM CVn stars, an `optimistic' and a `pessimistic' model, each taking into account the two dominant channels. In the optimistic model, there is a strong tidal coupling between white dwarf binaries in the WD channel, which helps to stabilize the mass transfer by feeding back angular momentum accreted by the primary star to the orbit. Furthermore, carbon-oxygen (CO) white dwarfs accreting from a He star are not easily destroyed by edge-lit detonations (ELDs) in this model. In the pessimistic model we assume no tidal coupling between white dwarfs, causing many to merge and few to survive as stably accreting AM CVn stars. In this model we furthermore include a favourable scenario for ELDs to destroy AM CVn progenitors in the He-star channel, although this has only a limited effect on the number of AM CVn stars emerging from this channel. See \citet{nelemans,nelemans2004} for a more quantitative description of the two scenarios.

In order to allow for the possibility that one of these two formation channels is shut off completely for some (as yet unknown) reason, we furthermore model optimistic and pessimistic populations for each of these channels individually.

All these population synthesis models include a more or less realistic Galactic model as in \citet{nelemans2004}, which assumes a star formation rate that depends on position and time as modelled by \citet{boissier}. All the stars that are formed are evolved using the latest version of the \texttt{SeBa} evolution code \citep{seba,nelemans,nelemans2004}. The resulting population of AM CVn stars at a distance $R=8.5$\,kpc from the Galactic centre and at time $t=13.5$\,Gyr is analysed. The only change we make to the model is that we assume a Galactic scale height of 300 pc for the AM CVn stars as in \citet{roelofshst}, instead of the 200 pc used in \citet{nelemans2004}. This is more appropriate for a population that is, on average, several Gyr old.

\subsection{Practical considerations}

The aforementioned parametrization of effective temperature with orbital period $P$, from which we derive the colours as a function of orbital period, is taken to be
\beq
T(P) = 19,000\,\mathrm{K}\times e^{-0.025\left(P-P_\mathrm{min}\right)}
\label{T_P}
\eeq
based on figure 2 of \citet{bildsten}, with $P$ in minutes and $P_\mathrm{min}=30$\,min as before. From the same figure we parametrize the absolute magnitude as a function of orbital period as
\beq
M_g(P) = 10.5 + 0.075\left(P-P_\mathrm{min}\right)
\label{Mg_P}
\eeq
with an estimated root-mean-square scatter of 0.5 magnitude. There are two emission-line AM CVn stars for which there is a parallax measurement: GP Com and V396 Hya. GP Com at $P=46$\,min has absolute magnitude $M_V=11.5$ \citep{roelofshst} and V396 Hya at $P=65$\,min has $M_V=13.3$ (J. Thorstensen, as quoted by \citealt{roelofshst}). Both match well with the above parametrization.

\begin{figure}
\centering
\includegraphics[angle=270,width=84mm]{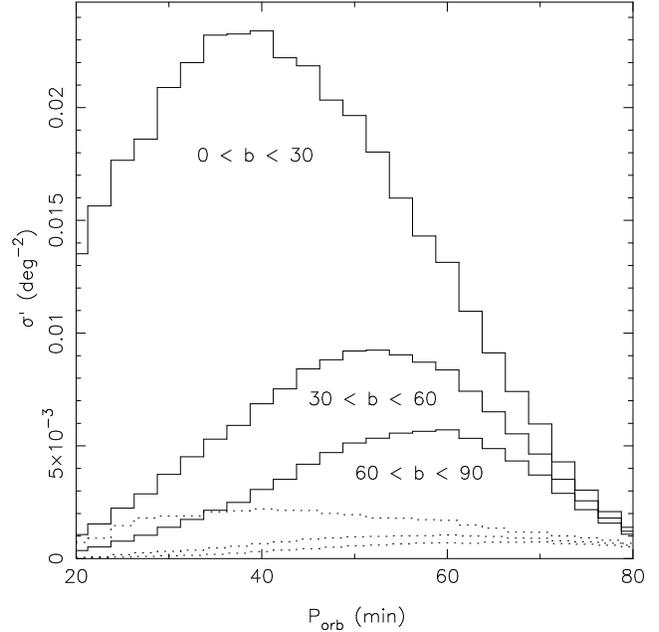}
\caption{Modelled surface density $\sigma'$ (per square degree) of AM CVn stars as a function of orbital period, down to $g=21$, for three Galactic latitude ranges. The solid line shows the `optimistic' model, the dotted line the corresponding `pessimistic' model.}
\label{population_b}
\end{figure}

For the completeness $C$ as well as for the model population $\rho'$ we have to construct a three-dimensional grid of values as a function of $P$, $b$ and $d$. Since the modelled space density $\rho'$ will (to a good approximation) only be a function of $P$ and Galactic height $z=d\sin(b)$, this reduces to a two-dimensional grid in $P$ and $z$. We bin the modelled systems with a resolution of 5 minutes in $P$ and 50 pc in $z$, from which we then fetch a population density at arbitrary $P$ and $z$ via bilinear interpolation. To give an idea of the model population, figure \ref{population_b} shows the number of systems as a function of orbital period $P$ and Galactic latitude $b$, integrated up to $g_\mathrm{max}=21$. At high Galactic latitudes, the population is suppressed relative to the low-latitude population as an effect of the population scale height. This suppression is stronger at short orbital periods due to the systems' higher brightness at short periods, shifting the peak in the distribution to longer-period, fainter (and thus closer) systems. We see that above $b\simeq 30^\circ$, which is where the sky coverage of the SDSS-I is concentrated, the effect of Galactic latitude on the expected population is modest.

\begin{figure}
\centering
\includegraphics[angle=270,width=84mm]{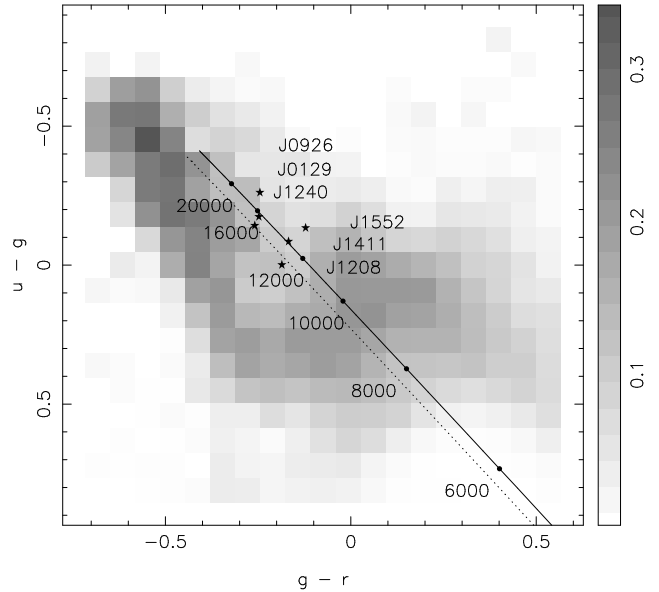}
\caption{Completeness of spectroscopic follow-up (grey-scale) as a function of $u-g$ and $g-r$, to a limiting magnitude $g=21$. Stars indicate the known AM CVns from SDSS-I. The solid line is the modelled cooling track for the AM CVns, closely following the blackbody cooling track (dotted line). Black dots indicate the `effective temperature' as goes into eq.\ \ref{T_P}.}
\label{completeness_colours}
\end{figure}

\begin{figure}
\centering
\includegraphics[angle=270,width=84mm]{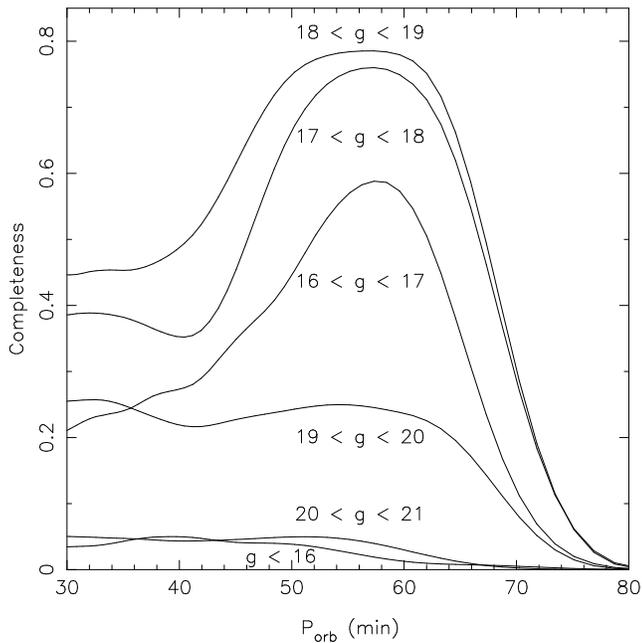}
\caption{Completeness of spectroscopic follow-up $C(P,g)$ as a function of the modelled orbital period $P$ and for different intervals in apparent magnitude $g$.}
\label{completeness_P}
\end{figure}

The completeness of spectroscopic follow-up $C(P,b,d)$ turns out to be insensitive to the Galactic latitude $b$ over the area covered by the SDSS-I and we thus determine the completeness only as a function of colour, $u-g$ and $g-r$, and apparent magnitude, $g$. Colour and apparent magnitude are again linked to orbital period and distance according to eqs. \ref{T_P} and \ref{Mg_P}. See figure \ref{completeness_colours} for the completeness as a function of colour, integrated up to $g_\mathrm{max}=21$. To the lower left of the blackbody track in this figure we see the targeted DA white dwarfs, which deviate from blackbodies due to the hydrogen absorption in their atmospheres, while the spectroscopic follow-up to the right of the blackbody track is mostly targeted at quasars. The six AM CVn stars from the SDSS-I are seen to lie close to the blackbody curve; we model their cooling track parallel to but very slightly offset from it, so that it runs through the middle of the bunch of observed systems. Figure \ref{completeness_P} shows the completeness along the modelled cooling track as a function of apparent magnitude $g$. It is seen to depend strongly on apparent magnitude, and on colour in the `cool' blackbody regime where the colours approach the bulk of the main sequence at $P\approx 70$\,min. We construct a two-dimensional grid for the completeness $C$ in intervals of 0.5 magnitude in $g$ and 5 minutes in orbital period $P$ from which we again get $C$ at arbitrary $P$ and $g$ via bilinear interpolation. Since the SDSS has five colour bands, $ugriz$, the completeness of spectroscopic follow-up is not uniquely defined for each point in $(u-g,g-r)$ space, but may in principle depend on the $iz$ colours. In the hot blackbody regime where the AM CVn stars are, however, it is unlikely that the $iz$ colours are of much influence. We verify this by determining the completeness as a function of $r-i$ and $i-z$ for the objects in a strip along the modelled cooling track from about 20,000--10,000\,K, which includes the 6 AM CVn stars found in the SDSS. It turns out to be a very flat distribution; in other words the completeness along the cooling track in $(u-g,g-r)$ is only weakly dependent on the $(r-i,i-z)$ colours of the corresponding objects.

Finally, we need to determine a suitable value $g_\mathrm{max}$ to delimit a well-defined sample of AM CVn stars. We choose $g_\mathrm{max}=21$ for two reasons: first, the completeness of spectroscopic follow-up in the SDSS plummets between $g=20.5$ and $g=21$ (see also figure \ref{completeness_P}), and second, up to this brightness limit we can still be confident that our efficiency for identifying emission-line AM CVn stars in the SDSS spectroscopic database is (nearly) 100\%. All six AM CVn stars identified in the spectroscopic database have $g<21$.

\subsection{Results}

\begin{table*}
\begin{center}
\begin{tabular}{l r r r r r r r}
\hline
Model		              &Modelled number          &Total in SDSS-I        &Modelled $\rho'_0$     &Observed $\rho_0$      &Observed $\Sigma_0$	&Per solar mass\\
                              &($N'_\mathrm{spec}$)     &($N_\mathrm{phot}$)    &(pc$^{-3}$)            &(pc$^{-3}$)            &(pc$^{-2}$)		&($M_\odot^{-1}$)\\
\hline
\hline
Optimistic                    &106                      &53                     &$2.7\times 10^{-5}$    &$1.5\times 10^{-6}$    &$4.5\times 10^{-4}$	&$2.2\times 10^{-5}$\\
Pessimistic                   &12                       &66                     &$6.1\times 10^{-6}$    &$3.1\times 10^{-6}$    &$9.4\times 10^{-4}$	&$4.5\times 10^{-5}$\\[1ex]

He star only, opt.            &16                       &68                     &$9.0\times 10^{-6}$    &$3.4\times 10^{-6}$    &$1.0\times 10^{-3}$	&$4.9\times 10^{-5}$\\
He star only, pess.           &11                       &67                     &$5.9\times 10^{-6}$    &$3.3\times 10^{-6}$    &$9.8\times 10^{-4}$	&$4.7\times 10^{-5}$\\
WD only, opt.                 &90                       &51                     &$1.8\times 10^{-5}$    &$1.2\times 10^{-6}$    &$3.5\times 10^{-4}$	&$1.7\times 10^{-5}$\\
WD only, pess.                &0.82                     &55                     &$2.2\times 10^{-7}$    &$1.6\times 10^{-6}$    &$4.6\times 10^{-4}$	&$2.3\times 10^{-5}$\\
\hline
\end{tabular}
\caption{Observed space densities of AM CVn stars for different assumptions regarding their populations. `Optimistic' and `pessimistic' models from \citet{nelemans} with the Galactic model of \citet{nelemans2004}. The observed local space density $\rho_0$ is obtained by multiplying the modelled $\rho'_0$ by $N_\mathrm{spec}/N'_\mathrm{spec}$ where $N_\mathrm{spec}=6$. The total $N_\mathrm{phot}$ is the number of emission-line AM CVn stars that should be present in the SDSS-I photometry down to $g_\mathrm{max}=21$. The measured local surface density $\Sigma_0$ projected onto the Galactic plane, in pc$^{-2}$, provides a scale-height-independent space density. The number of AM CVns per unit mass (in $M_\odot^{-1}$) normalizes this number to the present-day, local stellar mass in the Galactic disk \citep{chabrier}. The observed numbers of AM CVns per pc$^2$ and per $M_\odot$ are accurate to an estimated factor of 2. The observed $\rho_0$ in addition scales inversely with the assumed scale height; the numbers given are for a scale height of 300 pc.}
\label{densities}
\end{center}
\end{table*}

Table \ref{densities} shows the expected number of emission-line AM CVn stars in the SDSS spectroscopic database based on the different population synthesis models. From this we obtain the `observed' local space density $\rho_0$ by multiplying the modelled $\rho'_0$ with $N_\mathrm{spec}/N'_\mathrm{spec}$ (eq.\ \ref{transform}), where $N_\mathrm{spec}=6$. We see that all models predict more systems than are observed, except for the pessimistic model in which additionally there are no AM CVns from the He-star channel. Although the space densities in the different models vary by more than two orders of magnitude, the observed space density based on these input populations varies only by a factor of 3, from $\rho_0=1-3\times 10^{-6}$\,pc$^{-3}$. As mentioned in section \ref{recent}, this is because the variation in the total number of systems produced in the different models is large whereas the variation in the orbital period distribution is quite small, and the latter is what causes a variation in the calibrated (observed) space density.

The uncertainties in the values in table \ref{densities} are dominated by the Poisson statistics of the observed sample ($N_\mathrm{spec}=6$) and the uncertainties in the parametrization of the temperature and absolute magnitude with orbital period (eqs.\ \ref{T_P} and \ref{Mg_P}). The small sample of 6 systems contributes an intrinsic uncertainty of order $6^{-1/2}\approx 40\%$ to the measured values in table \ref{densities}. We shift the parametrization of absolute magnitude with orbital period (eq.\ \ref{Mg_P}) up and down by 0.3 magnitudes as an estimated uncertainty; this yields a variation of 32\% in the observed space densities. Similarly we vary the effective temperature as a function of orbital period with $\pm 20\%$ and find a variation of 3--10\% in the resulting space densities. Adding up all the numbers, we thus estimate a total uncertainty in the derived space densities that is close to a factor of 2 for each of the model populations.

Note that the pessimistic model from \citet{nelemans}, when calibrated with the observed systems from the SDSS, actually yields a \emph{larger} space density than the optimistic model. This is due to the fact that the pessimistic model is dominated by AM CVns from the He-star channel, which is on average a `faster' channel that follows the star formation more closely. These systems are thus on average older and at longer orbital periods (and therefore fainter), causing the average completeness of spectroscopic follow-up to be lower and the resulting space density to be higher.

\section{Discussion}

\subsection{Previous space density estimates}

\citet{warnerbook} estimates a local space density $\rho_0\sim 2\times 10^{-6}$\,pc$^{-3}$, identical to our result. Interestingly though, Warner arrives at this value for the wrong reasons: his estimate is essentially based on the short-period systems, and fails to take into account the large population of long-period systems that should accompany them (as mentioned before, the fraction of short-period systems is expected to be $\leq 2\%$). But this is exactly offset by the assumption that the short-period systems have absolute magnitudes $M_V=9.5$, which from \emph{HST} parallax measurements we now know is actually $M_V\simeq 6$ \citep{roelofshst}.

The estimate for the space density based on the short-period AM CVn stars was recently updated in \citet{roelofshst} based on the updated absolute magnitudes from parallax measurements, giving $\rho_0\sim 1\times 10^{-6}/q$~pc$^{-3}$, where $q$ is the known fraction of short-period systems with apparent magnitude $m_V<14.5$. The space density derived here suggests that the unknown fraction of bright, short-period systems is rather small. Alternatively, the comparable numbers could point towards a destruction of systems during their evolution from short to long orbital periods if one assumes that there should in fact be a significant population of short-period systems that have not yet been discovered. After all, the known AM CVns have been discovered serendipitously and predominantly at high Galactic latitudes, while the majority of short-period systems with $m_V<14.5$ should in fact reside at low latitude, given their $M_V\simeq 6$.

\subsection{Population synthesis results}

When comparing the observed space density to the models, one has to keep in mind that there are two different factors of uncertainty in the models: (a) the formation and evolution of AM CVn stars, and (b) the Galactic model containing the position- and time-dependent star formation history. This makes it more difficult to draw conclusions about one of the two based on the observed space densities.

Fortunately the uncertainties due to the Galactic model are expected to be limited. We are looking at the local population of AM CVn stars, and what matters thus is the local star formation history, which is calibrated observationally with for instance the local stellar density, giving the integrated star formation history (see \citealt{boissier}). Inclusion of a star formation history that is a function of both time and distance from the Galactic centre, versus one where the time-dependence is independent of position, has lowered the local space density of AM CVn stars by a factor 2.7 \citep{nelemans,nelemans2004}. This is due to the fact that the AM CVn star progeny are relatively old populations; the delay in the star formation at $R=8.5$\,kpc relative to the centre of the Galaxy reduces the number of systems in the solar neighbourhood that have already evolved to become AM CVn stars. We estimate that the remaining room for variations in the modelled space density of AM CVn stars due to the local star formation history is less than a factor of 2.

\begin{figure}
\centering
\includegraphics[angle=270,width=84mm]{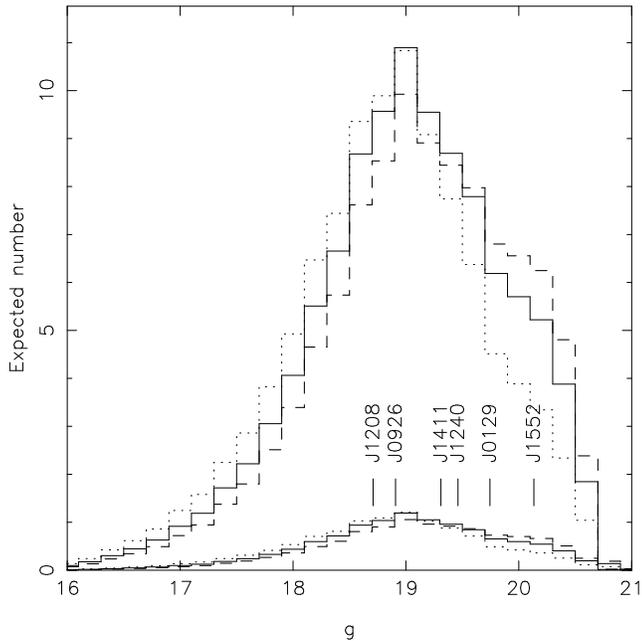}
\caption{Modelled distribution of spectroscopically identified AM CVn stars in SDSS-I as a function of apparent magnitude $g$. The dotted, solid and dashed lines represent model population scale heights of 200, 300 and 400 pc, respectively. The upper set of lines are for the optimistic model; the pessimistic model (lower lines) gives virtually identical results but scaled downwards.}
\label{expected_g}
\end{figure}

What could further be of influence is the assumed scale height $h_z=300$\,pc of the AM CVn stars above the Galactic plane. Figure \ref{expected_g} shows the modelled and observed distributions in apparent magnitude $g$ for different scale heights of $h_z=200,300$ and $400$\,pc. The effect on the total number of expected systems in the SDSS-I is clearly very small, indicating that the discrepancy between the observed and the modelled space densities is not due to the assumed scale height of the model population. Note also that the larger scale heights appear to be slightly (though not significantly) preferred based on the observed sample. Although the ratio between the modelled and the observed space densities is insensitive to the assumed scale height, the absolute numbers are: if assuming a scale height of 400 pc instead of 300 pc, both numbers need to be multiplied by approximately $3/4$. In order to provide a scale-height-independent number, which is essentially what we obtain from the SDSS-I since we are looking out to several scale heights for the average AM CVn star, we provide in table \ref{densities} the surface density $\Sigma$ of AM CVns when we project them all onto the Galactic plane, in units of pc$^{-2}$. Adopting a local surface density of $21M_\odot/\mathrm{pc}^{-2}$ for stellar matter in the Galactic disk \citep{chabrier}, we also give the local number of AM CVns per solar mass, in the last column of table \ref{densities}.

We conclude that a substantial part of the overoptimism in the `optimistic' model (which yields a factor 18 higher space density than observed, see table \ref{densities}) is due to the evolutionary model of the AM CVn stars. The easiest explanation is that the optimistic model for the WD channel is far too optimistic: mass transfer between white dwarfs upon Roche-lobe overflow cannot be stabilized in the majority of cases, but rather leads to merger events \citep{nelemans,masstransfer}. The pessimistic model for the WD channel however, which assumes no stabilizing effect on the mass transfer from feedback of angular momentum due to tidal coupling between white dwarfs, yields predictions for the space density that are too low. This means that there is either a contribution from the He-star channel, which by itself produces modelled space densities that are compatible with the observed value, or that there may be some stabilizing effect due to tidal coupling in the WD channel.

The two most viable scenarios in the framework of the current models are therefore:
\begin{enumerate}
\item There is a mild tidal coupling between white dwarfs, which helps to stabilize the mass transfer in up to 10\% of all cases in which this could work, while the He-star channel is inactive;
\item The He-star channel is active and there is no efficient tidal coupling between white dwarfs in the WD channel, causing the WD channel to be of little influence.
\end{enumerate}

\subsection{Implications for AM CVn star evolution}

\begin{figure}
\centering
\includegraphics[angle=270,width=84mm]{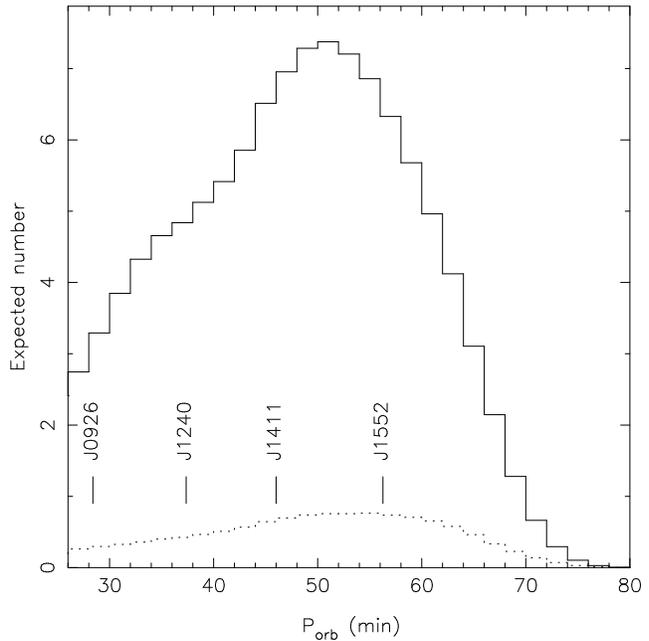}
\caption{Modelled distribution of spectroscopically identified AM CVn stars in SDSS-I as a function of orbital period, along our modelled cooling track. The 4 Sloan AM CVns for which a \emph{measured} orbital period exists are indicated in the figure. The solid and dotted lines are the optimistic and pessimistic models, respectively.}
\label{expected_P}
\end{figure}

If we look outside the box of the current models, the relatively low number of observed emission-line AM CVn stars may most readily be explained by assuming that long-period AM CVn stars do not always exhibit the helium emission lines by which they were identified from the SDSS-I. It has been suggested from donor star model calculations \citep{deloye} that in AM CVn stars with hot, semi-degenerate donors, these donors may cool and contract, so that the mass transfer rate goes down substantially and, depending on how the timescales work out, they might even become detached. This would eliminate them from the sample studied here. Essentially, such semi-degenerate donors cool down when their (large) thermal timescale \citep{savonije,tf} finally catches up with the (rapidly increasing) timescale of angular momentum loss from the system due to the emission of gravitational waves. Calculations of the thermal evolution of the donor stars in AM CVns (Deloye in preparation; private communication) indicate that this could happen at $P\approx 45$ min. Observationally, it has recently been found that the donor stars in short-period AM CVns ($P<30$\,min) appear to be mostly semi-degenerate \citep{roelofshst,nasser}; it would thus appear that this scenario is possible.

Observationally, this would become clear as a fall-off of the population at $P>45$\,min or a pile-up of systems around $P\approx45$\,min, depending on whether they become fully detached. The current sample of six, of which orbital periods have only been measured for four, at present prevents us from drawing conclusions. The fact that AM CVns at orbital periods well above 45 minutes have been observed, indicates that at least not all of them become detached indefinitely. In addition, even if this were the case, it is clear from figure \ref{completeness_colours} that this could at most have caused a suppression of the population by a factor of 3, based on the fraction of observed systems that is expected to be at orbital periods $P>45$\,min.


The He-star models take into account optimistic and pessimistic estimates for the destruction of AM CVn stars by edge-lit detonations of the accreting CO white dwarf \citep{nelemans}. Although recent studies question the existence of such ELDs altogether (e.g.\ \citealt{yoon}), their possible non-existence is not very relevant for the AM CVn population since it is of relatively little influence on the number of AM CVn stars produced in the He-star channel (see table \ref{densities}). What could be of influence is the suggestion that the He-star channel may hardly work at all due to the poor efficiency for ejecting the common envelope in such a configuration. This is thought to be due to the hydrodynamics of the relatively weak density gradient between the He-burning core and its hydrogen mantle, which together form the He-star progenitor (see \citet{sandquist} for a study of how this may prevent the formation of low-mass ($<0.2\,M_\odot$) He WDs from a common envelope).

In a variation on the ELD theme, \citet{.Ia} have recently shown that a substantial fraction of AM CVns, and possibly all of them, could host a strong thermonuclear event due to the ignition of a sufficiently thick layer of accreted helium on the accretor's CO core. This could destroy the binary and thus remove AM CVns from the population \emph{after} they have started their evolution towards longer orbital periods as stable mass-transferring binaries.

In the WD channel, the effect of the unknown tidal coupling between the accretor and the donor on their survival as stable mass-transferring white dwarfs \citep{nelemans,masstransfer} has been taken into account by modelling an optimistic (perfect coupling) and a pessimistic (no coupling) scenario. In the latter case, the WD channel is so much suppressed due to the mass transfer becoming unstable that the observed space density is too high for this channel to be the only contributor. An additional, relatively unstudied effect is the possible ignition of He-core white dwarfs upon accretion of a substantial amount of helium from their helium-white-dwarf donors, either through a series of inward-moving helium flashes \citep{saio} or via an inward-moving burning front (Bildsten in preparation; private communication). Such He+He WD binaries represent up to 20\% of the AM CVn star progenitors in the WD channel for the optimistic model (but note again the arguments by \citet{sandquist} against the formation of low-mass He WDs in such binaries); the possible non-existence of AM CVns descending from He+He WD binaries should therefore be of limited influence. In the pessimistic model, only WD binaries of extreme mass ratio survive the initial mass transfer, leaving no AM CVn stars from He+He white dwarf binaries anyway.

With the lower space densities reported in this paper, we note that the evolved-CV channel for the formation of AM CVn stars \citep{podsi}, which was previously estimated to contribute at the $<2\%$ level (\citealt{nelemans2004}), might become important. However, AM CVns from this channel should either (a) pile up more strongly at long orbital periods due to the fact that most systems are not sufficiently hydrogen-depleted to shift their orbital period minima all the way down to $\sim$10 minutes; or (b) look like `ordinary' CVs due to their remaining hydrogen, implying that there should be many CVs with orbital periods well below one hour. Neither of that appears to be observed. Nevertheless the observational challenge of finding an AM CVn star with traces of hydrogen in its accretion disc remains.

\subsection{Implications for the AM CVn stars as \emph{LISA} sources}

AM CVn stars are thought to be among the most important sources for future space-borne gravitational-wave detectors such as \emph{LISA} (e.g.\ \citealt{nelemans2004}). In addition, it is currently the only class of sources for which known, resolvable sources exist \citep{stroeer2005,stroeer2006,roelofshst}. Current estimates for the number of AM CVn stars resolvable with a \emph{LISA}-type gravitational-wave detector are based on the `optimistic' model for their formation and evolution \citep{nelemans2004}. The results presented here suggest that these estimates should be lowered. As discussed in the previous section, a loss of systems during their evolution from the most sensitive \emph{LISA} band ($P\lesssim 20$\,min) to the emission-line regime at $P>30$\,min could cause us to underestimate the short-period population. Nevertheless it would seem reasonable, based on the results shown in table \ref{densities}, to lower the estimate of $\sim$$11,000$ Galactic AM CVn stars that can be resolved with \emph{LISA} to about $\sim$$1,000$, for a mission duration of one year \citep{nelemans2004}.

\section{Conclusion}

We have investigated the population of helium-emission-line AM CVn stars based on the sample of six new systems from the Sloan Digital Sky Survey. We have compared the observed sample to predicted samples as obtained from population synthesis models. For different model populations we have derived \emph{observed} local space densities of $1-3\times 10^{-6}$\,pc$^{-3}$. In addition to the variations in observed space density for the different models, we estimate uncertainties of a factor of 2 due to Poisson noise and (assumed) uncertainties in the colours and absolute magnitudes of the emission-line AM CVn stars. In the standard evolutionary picture, where all AM CVn stars evolve to become emission-line systems at orbital periods above $\sim$30 minutes, this corresponds to a space density of the entire population that is very nearly identical to this value (within 2\%). A further refinement of the space density would require additional follow-up spectroscopy of photometric candidates in the SDSS, to increase the spectroscopic sample and bring down the Poisson noise. Since there are relatively few other types of objects with approximate hot blackbody colours (mainly single DB white dwarfs), this is feasible. Further orbital period measurements would be needed to reduce the uncertainty in the colour evolution and orbital period distribution of the AM CVn population.

The observed space density presented here is lower than expected from the population synthesis models, which predict $6\times 10^{-6}$\,pc$^{-3}$ for the pessimistic models to $3\times 10^{-5}$\,pc$^{-3}$ for the optimistic models. At least one of the proposed dominant formation channels, the double-degenerate WD channel, has to be suppressed by at least an order of magnitude compared to the optimistic models, which assume that the mass transfer between white dwarfs can be stabilized in many cases due to a strong tidal coupling of spin and orbital angular momentum. A significant effect of tidal coupling on the survival rate of AM CVns from the WD channel is possible only if the second dominant formation channel, the single-degenerate He-star channel, is also severely suppressed.

We have presented an inventory of ideas, based on current theory, of how the formation channels may be suppressed relative to the models we have used. Most of these (detachment of long-period AM CVns, ignition of He+He WD accretors) have only a limited effect, but inefficient common-envelope ejection in the He-star channel could potentially be effective in shutting down the He-star channel completely.

\section{Acknowledgments}

It is a pleasure to thank Jim Liebert and Chris Deloye for stimulating discussions. GHAR and PJG were supported by NWO VIDI grant 639.042.201 to P.J. Groot. GN was supported by NWO VENI grant 639.041.405 to G. Nelemans.

Funding for the SDSS has been provided by the Alfred P. Sloan Foundation, the Participating Institutions, the National Science Foundation, the U.S. Department of Energy, the National Aeronautics and Space Administration, the Japanese Monbukagakusho, the Max Planck Society, and the Higher Education Funding Council for England. The SDSS is managed by the Astrophysical Research Consortium for the Participating Institutions. The Participating Institutions are the American Museum of Natural History, Astrophysical Institute Potsdam, University of Basel, Cambridge University, Case Western Reserve University, University of Chicago, Drexel University, Fermilab, the Institute for Advanced Study, the Japan Participation Group, Johns Hopkins University, the Joint Institute for Nuclear Astrophysics, the Kavli Institute for Particle Astrophysics and Cosmology, the Korean Scientist Group, the Chinese Academy of Sciences (LAMOST), Los Alamos National Laboratory, the Max-Planck-Institute for Astronomy (MPIA), the Max-Planck-Institute for Astrophysics (MPA), New Mexico State University, Ohio State University, University of Pittsburgh, University of Portsmouth, Princeton University, the United States Naval Observatory, and the University of Washington.

\end{document}